\documentstyle[twoside]{article}

\catcode`\@=11
\long\def\@makefntext#1{
\protect\noindent \hbox to 3.2pt {\hskip-.9pt  
$^{{\eightrm\@thefnmark}}$\hfil}#1\hfill}               

\def\@makefnmark{\hbox to 0pt{$^{\@thefnmark}$\hss}}    
        
\def\ps@myheadings{\let\@mkboth\@gobbletwo
\def\@oddhead{\hbox{}
\rightmark\hfil\eightrm\thepage}   
\def\@oddfoot{}\def\@evenhead{\eightrm\thepage\hfil
\leftmark\hbox{}}\def\@evenfoot{}
\def\sectionmark##1{}\def\subsectionmark##1{}}

\oddsidemargin=\evensidemargin
\newcounter{sectionc}\newcounter{subsectionc}\newcounter{subsubsectionc}
\renewcommand{\section}[1] {\vspace{12pt}\addtocounter{sectionc}{1} 
\setcounter{subsectionc}{0}\setcounter{subsubsectionc}{0}\noindent 
        {\tenbf\thesectionc. #1}\par\vspace{5pt}}
\renewcommand{\subsection}[1] {\vspace{12pt}\addtocounter{subsectionc}{1} 
        \setcounter{subsubsectionc}{0}\noindent 
        {\bf\thesectionc.\thesubsectionc. {\kern1pt \bfit #1}}\par\vspace{5pt}}
\renewcommand{\subsubsection}[1] {\vspace{12pt}\addtocounter{subsubsectionc}{1}
        \noindent{\tenrm\thesectionc.\thesubsectionc.\thesubsubsectionc.
        {\kern1pt \tenit #1}}\par\vspace{5pt}}
\newcommand{\nonumsection}[1] {\vspace{12pt}\noindent{\tenbf #1}
        \par\vspace{5pt}}

\newcounter{appendixc}
\newcounter{subappendixc}[appendixc]
\newcounter{subsubappendixc}[subappendixc]
\renewcommand{\thesubappendixc}{\Alph{appendixc}.\arabic{subappendixc}}
\renewcommand{\thesubsubappendixc}
        {\Alph{appendixc}.\arabic{subappendixc}.\arabic{subsubappendixc}}

\renewcommand{\appendix}[1] {\vspace{12pt}
        \refstepcounter{appendixc}
        \setcounter{figure}{0}
        \setcounter{table}{0}
        \setcounter{lemma}{0}
        \setcounter{theorem}{0}
        \setcounter{corollary}{0}
        \setcounter{definition}{0}
        \setcounter{equation}{0}
        \renewcommand{\thefigure}{\Alph{appendixc}.\arabic{figure}}
        \renewcommand{\thetable}{\Alph{appendixc}.\arabic{table}}
        \renewcommand{\theappendixc}{\Alph{appendixc}}
        \renewcommand{\thelemma}{\Alph{appendixc}.\arabic{lemma}}
        \renewcommand{\thetheorem}{\Alph{appendixc}.\arabic{theorem}}
        \renewcommand{\thedefinition}{\Alph{appendixc}.\arabic{definition}}
        \renewcommand{\thecorollary}{\Alph{appendixc}.\arabic{corollary}}
        \renewcommand{\theequation}{\Alph{appendixc}.\arabic{equation}}
        \noindent{\tenbf Appendix \theappendixc #1}\par\vspace{5pt}}
\newcommand{\subappendix}[1] {\vspace{12pt}
        \refstepcounter{subappendixc}
        \noindent{\bf Appendix \thesubappendixc. {\kern1pt \bfit #1}}
        \par\vspace{5pt}}
\newcommand{\subsubappendix}[1] {\vspace{12pt}
        \refstepcounter{subsubappendixc}
        \noindent{\rm Appendix \thesubsubappendixc. {\kern1pt \tenit #1}}
        \par\vspace{5pt}}

\newcommand{\textlineskip}{\baselineskip=13pt}
\newcommand{\smalllineskip}{\baselineskip=10pt}

\def\eightcirc{
\begin{picture}(0,0)
\put(4.4,1.8){\circle{6.5}}
\end{picture}}
\def\eightcopyright{\eightcirc\kern2.7pt\hbox{\eightrm c}}

\def\abstracts#1#2#3{{
        \centering{\begin{minipage}{4.5in}\baselineskip=10pt\footnotesize
        \parindent=0pt #1\par 
        \parindent=15pt #2\par
        \parindent=15pt #3
        \end{minipage}}\par}} 

\renewenvironment{thebibliography}[1]
        {\frenchspacing
         \ninerm\baselineskip=11pt
         \begin{list}{\arabic{enumi}.}
        {\usecounter{enumi}\setlength{\parsep}{0pt}
         \setlength{\leftmargin 17pt}{\rightmargin 0pt}   
         \setlength{\itemsep}{0pt} \settowidth
        {\labelwidth}{#1.}\sloppy}}{\end{list}}

\newcounter{itemlistc}
\newcounter{romanlistc}
\newcounter{alphlistc}
\newcounter{arabiclistc}

\newcommand{\fcaption}[1]{
        \refstepcounter{figure}
        \setbox\@tempboxa = \hbox{\footnotesize Fig.~\thefigure. #1}
        \ifdim \wd\@tempboxa > 5in
           {\begin{center}
        \parbox{5in}{\footnotesize\smalllineskip Fig.~\thefigure. #1}
            \end{center}}
        \else
             {\begin{center}
             {\footnotesize Fig.~\thefigure. #1}
              \end{center}}
        \fi}

\newcommand{\tcaption}[1]{
        \refstepcounter{table}
        \setbox\@tempboxa = \hbox{\footnotesize Table~\thetable. #1}
        \ifdim \wd\@tempboxa > 5in
           {\begin{center}
        \parbox{5in}{\footnotesize\smalllineskip Table~\thetable. #1}
            \end{center}}
        \else
             {\begin{center}
             {\footnotesize Table~\thetable. #1}
              \end{center}}
        \fi}

\def\@citex[#1]#2{\if@filesw\immediate\write\@auxout
        {\string\citation{#2}}\fi
\def\@citea{}\@cite{\@for\@citeb:=#2\do
        {\@citea\def\@citea{,}\@ifundefined
        {b@\@citeb}{{\bf ?}\@warning
        {Citation `\@citeb' on page \thepage \space undefined}}
        {\csname b@\@citeb\endcsname}}}{#1}}

\newif\if@cghi
\def\cite{\@cghitrue\@ifnextchar [{\@tempswatrue
        \@citex}{\@tempswafalse\@citex[]}}
\def\citelow{\@cghifalse\@ifnextchar [{\@tempswatrue
        \@citex}{\@tempswafalse\@citex[]}}
\def\@cite#1#2{{$\null^{#1}$\if@tempswa\typeout
        {IJCGA warning: optional citation argument 
        ignored: `#2'} \fi}}

\def\pmb#1{\setbox0=\hbox{#1}
        \kern-.025em\copy0\kern-\wd0
        \kern.05em\copy0\kern-\wd0
        \kern-.025em\raise.0433em\box0}

\def\fnt#1#2{\footnotetext{\kern-.3em
        {$^{\mbox{\scriptsize #1}}$}{#2}}}

\def\fpage#1{\begingroup
\voffset=.3in
\thispagestyle{empty}\begin{table}[b]\centerline{\footnotesize #1}
        \end{table}\endgroup}

\def\runninghead#1#2{\pagestyle{myheadings}
\markboth{{\protect\footnotesize\it{\quad #1}}\hfill}
{\hfill{\protect\footnotesize\it{#2\quad}}}}
\font\tenrm=cmr10
\font\tenit=cmti10 
\font\tenbf=cmbx10
\font\bfit=cmbxti10 at 10pt
\font\ninerm=cmr9

\font\eightrm=cmr8

\def\qed{\hbox{${\vcenter{\vbox{                        
   \hrule height 0.4pt\hbox{\vrule width 0.4pt height 6pt
   \kern5pt\vrule width 0.4pt}\hrule height 0.4pt}}}$}}

\newcommand{\zp}[3]{Z.\ Phys.\ {\bf C#1} (19#2) #3}
\newcommand{\pl}[3]{Phys.\ Lett.\ {\bf B#1} (19#2) #3}
\newcommand{\plold}[3]{Phys.\ Lett.\ {\bf #1B} (19#2) #3}
\newcommand{\np}[3]{Nucl.\ Phys.\ {\bf B#1} (19#2) #3}
\newcommand{\prd}[3]{Phys.\ Rev.\ {\bf D#1} (19#2) #3}
\newcommand{\prl}[3]{Phys.\ Rev.\ Lett.\ {\bf #1} (19#2) #3}
\newcommand{\mpl}[3]{Mod.\ Phys.\ Lett.\ {\bf A#1} (19#2) #3}

\newcommand{\md}{\mbox{d}}
\def\simgt{\rlap{\lower 3.5 pt \hbox{$\mathchar \sim$}} \raise 1pt \hbox {$>$}}
\def\simlt{\rlap{\lower 3.5 pt \hbox{$\mathchar \sim$}} \raise 1pt \hbox {$<$}}

\newcommand{\beq}{\begin{equation}}
\newcommand{\eeq}{\end{equation}}
\newcommand{\bea}{\begin{eqnarray}}
\newcommand{\eea}{\end{eqnarray}}

\newcommand{\uone}{\mbox{$\underline{1}$}}
\newcommand{\ueight}{\mbox{$\underline{8}$}}
\begin{document}

\runninghead{Color-Singlet and Color-Octet Contributions to 
$J/\psi$ Photoproduction}  {Color-Singlet and Color-Octet 
Contributions to $J/\psi$ Photoproduction}

\normalsize\textlineskip
\thispagestyle{empty}
\setcounter{page}{1}

\begin{flushright}
DESY 96-201\\
September 1996
\end{flushright}


\vspace*{-1.5cm}

\vspace*{0.88truein}

\fpage{1}
\centerline{\large{\bf Color-Singlet and Color-Octet Contributions}}
\vspace*{0.035truein}
\centerline{\large{\bf to $J/\psi$ Photoproduction\footnote{Talk presented 
       at the Workshops `QED and QCD in Higher Orders', Rheinsberg, Germany, 
       April 21-26 and `Quarkonium Physics', Chicago, USA, June 13-15, 
       1996; to appear in the proceedings.}}}
\vspace*{0.27truein}
\centerline{\footnotesize MICHAEL KR\"AMER}
\vspace*{0.015truein}
\centerline{\footnotesize\it Deutsches Elektronen-Synchrotron DESY, 
                             D-22603 Hamburg, FRG}

\vspace*{0.21truein}
\abstracts{
  I discuss the impact of color-octet contributions and higher-order
  QCD corrections on the cross section for inelastic $J/\psi$ 
  photoproduction. The theoretical predictions are compared with recent 
  experimental data obtained at \mbox{HERA}.
}{}{}

\vspace*{1pt}\textlineskip      
\section{Introduction}         
\vspace*{-0.5pt}
\noindent
The production of heavy quarkonium states in high-energy collisions
provides an important tool to study the interplay between perturbative
and non-perturbative QCD dynamics. While the creation of heavy quarks
in a hard scattering process can be calculated in perturbative
QCD\cite{CSS86}, the subsequent transition to a physical bound state
introduces non-perturbative aspects. A rigorous framework for treating
quarkonium production and decays has recently been
developed.\cite{BBL95} The factorization approach is based on the use
of non-relativistic QCD\cite{CL86} (NRQCD) to separate the
short-distance parts from the long-distance matrix elements and
explicitly takes into account the complete structure of the quarkonium
Fock space. This formalism implies that so-called color-octet
processes, in which the heavy-quark antiquark pair is produced at
short distances in a color-octet state and subsequently evolves
non-perturbatively into a physical quarkonium, should contribute to
the cross section.  It has recently been argued\cite{TEV1,CL96} that
quarkonium production in hadronic collisions at the Tevatron can be
accounted for by including color-octet processes and by adjusting the
unknown long-distance color-octet matrix elements to fit the data.

In order to establish the phenomenological significance of the
color-octet mechanism it is necessary to identify color-octet
contributions in different production processes. Color-octet
production of $J/\psi$ particles has also been studied in the context
of $e^+e^-$ annihilation\cite{BC95}, $Z$ decays\cite{CKY95}, hadronic
collisions at fixed-target experiments\cite{fthad,BR} and $B$
decays\cite{KLS2}.  Here, I review the impact of color-octet
contributions and higher-order QCD corrections on the cross section
for $J/\psi$ photoproduction. The production of $J/\psi$ particles in
photon-proton collisions proceeds predominantly through photon-gluon
fusion. Elastic/diffractive mechanisms\cite{ELASTIC} can be eliminated
by measuring the $J/\psi$ energy spectrum, described by the scaling
variable $z = {p\cdot k_\psi}\, / \, {p\cdot k_\gamma}$, with $p,
k_{\psi,\gamma}$ being the momenta of the proton and $J/\psi$,
$\gamma$ particles, respectively. In the proton rest frame, $z$ is the
ratio of the $J/\psi$ to $\gamma$ energy, $z=E_{\psi}/E_\gamma$. For
elastic/diffractive events $z$ is close to one; a clean sample of
inelastic events can be obtained in the range $z\;\simlt\;0.9$.

According to the NRQCD factorization formalism , the inclusive cross
section for $J/\psi$ photoproduction can be expressed as a sum of
terms, each of which factors into a short-distance coefficient and a
long-distance matrix element:
\begin{equation}\label{eq_fac}
\mbox{d}\sigma(\gamma+g \to J/\psi +X) = 
\sum_n \mbox{d}\hat{\sigma}(\gamma+g \to c\bar{c}\, [n] + X)\, 
  \langle {\cal{O}}^{J/\psi}\,[n] \rangle 
\end{equation}
Here, $\mbox{d}\hat{\sigma}$ denotes the short-distance cross section
for producing an on-shell $c\bar{c}$-pair in a color, spin and
angular-momentum state labelled by $n$.  The NRQCD matrix elements
$\langle {\cal{O}}^{J/\psi} \, [n] \rangle \equiv \langle 0 |
{\cal{O}}^{J/\psi} \, [n] | 0 \rangle$ give the probability for a
$c\bar{c}$-pair in the state $n$ to form the $J/\psi$ particle. The
relative importance of the various terms in (\ref{eq_fac}) can be
estimated by using NRQCD velocity scaling rules.\cite{LMNMH92} For
$v\to 0$ ($v$ being the average velocity of the charm quark in the
$J/\psi$ rest frame) each of the NRQCD matrix elements scales with a
definite power of $v$ and the general expression (\ref{eq_fac}) can be
organized into an expansion in powers of $v^2$.

\vspace*{1pt}\textlineskip      
\section{Color-singlet contribution}        
\begin{figure}[t]

\vspace*{2.75cm}

\begin{picture}(7,7)
\includegraphics{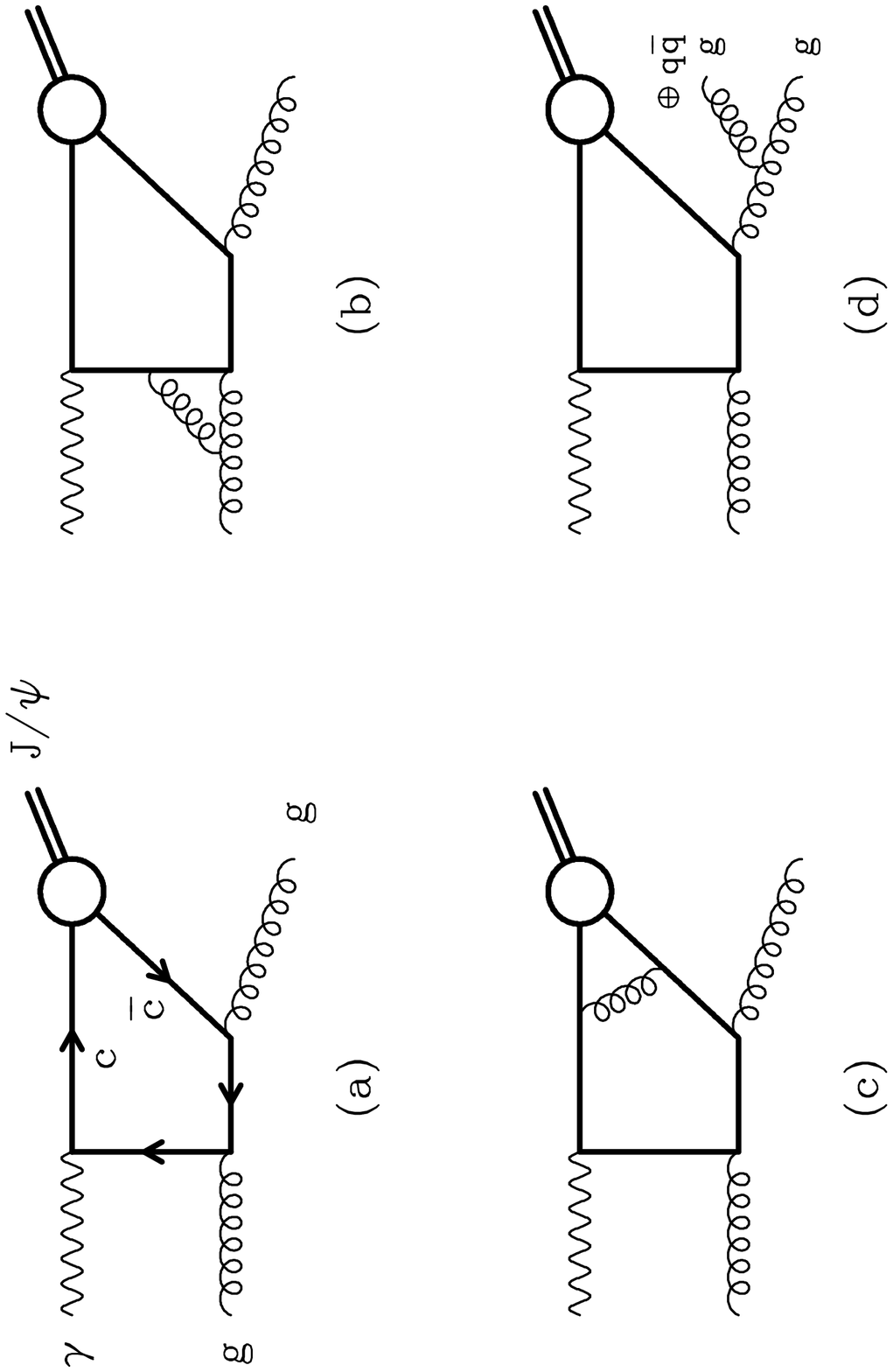}
\end{picture}

\vspace*{5.25cm}

\begin{picture}(7,7)
\includegraphics{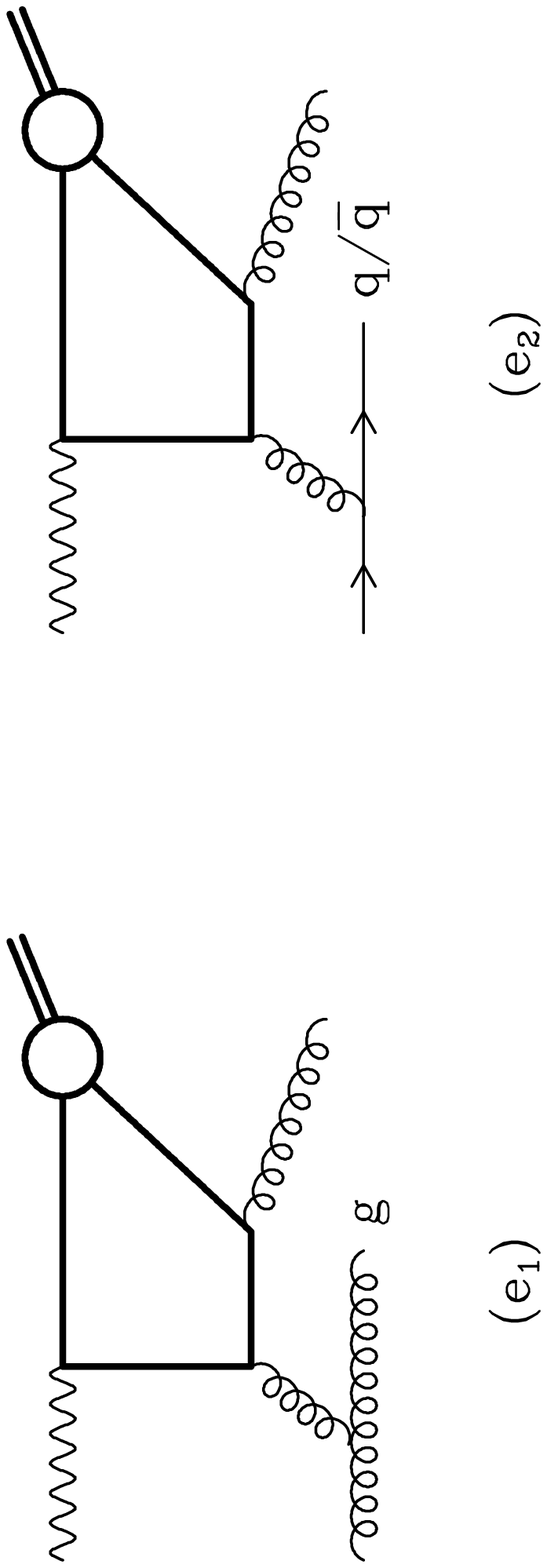}
\end{picture}

\vspace*{0.45cm}

\fcaption{\label{fig_1} Generic diagrams for $J/\psi$ 
  photoproduction via the color-singlet channel: (a) leading order
  contribution; (b) vertex corrections; (c) box diagrams; (d)
  splitting of the final state gluon into gluon or light
  quark-antiquark pairs; (e) diagrams renormalizing the initial-state
  parton densities.}

\vspace*{-5mm}

\end{figure}

\vspace*{-0.5pt}
\noindent
At leading order in $v^2$, eq.(\ref{eq_fac}) reduces to the standard
factorization formula of the color-singlet model\cite{CS}. The
short-distance cross section is given by the subprocess
\begin{equation}\label{eq_cs}
\gamma + g \to c\bar{c}\, [\mbox{$\underline{1}$},{}^3S_{1}] + g
\end{equation}
shown in Fig.\ref{fig_1}a, with $c\bar{c}$ in a color-singlet state
(denoted by \mbox{$\underline{1}$}), zero relative velocity, and
spin/angular-momentum quantum numbers $^{2S+1}L_J = {}^3S_1$.  Up to
corrections of ${\cal{O}}(v^4)$, the color-singlet NRQCD matrix
element is related to the $J/\psi$ wave function at the origin through
$\langle {\cal{O}}^{J/\psi}\,[\uone,{}^3S_{1}] \rangle \approx
(9/2\pi)|\varphi(0)|^2$ and can be extracted from the measurement of
the $J/\psi$ leptonic decay width or calculated within potential
models. Relativistic corrections due to the motion of the charm quarks
in the $J/\psi$ bound state enhance the large-$z$ region, but can be
neglected in the inelastic domain.\cite{REL} The calculation of the
higher-order perturbative QCD corrections to the short-distance cross
section (\ref{eq_cs}) has been performed recently.\cite{KZSZ94,MK95}
Generic diagrams which build up the cross section in next-to-leading
order (NLO) are depicted in Fig.\ref{fig_1}. Besides the usual
self-energy diagrams and vertex corrections for photons and gluons
(b), one encounters box diagrams (c), the splitting of the final-state
gluon into gluon and light quark-antiquark pairs, as well as diagrams
renormalizing the initial state parton densities (e). 
Inclusion of the NLO corrections reduces the scale dependence of the
theoretical prediction and increases the cross section significantly,
depending in detail on the $\gamma p$ energy and the choice of
parameters.\cite{MK95} Details of the calculation and a comprehensive
analysis of total cross sections and differential distributions for
the energy range of the fixed-target experiments and for $J/\psi$
photoproduction at \mbox{HERA} can be found elsewhere.\cite{MK95}

\newpage

\vspace*{1pt}\textlineskip      
\section{Color-octet contributions}
\begin{figure}[t]

\vspace*{2.5cm}

\begin{picture}(7,7)
\includegraphics{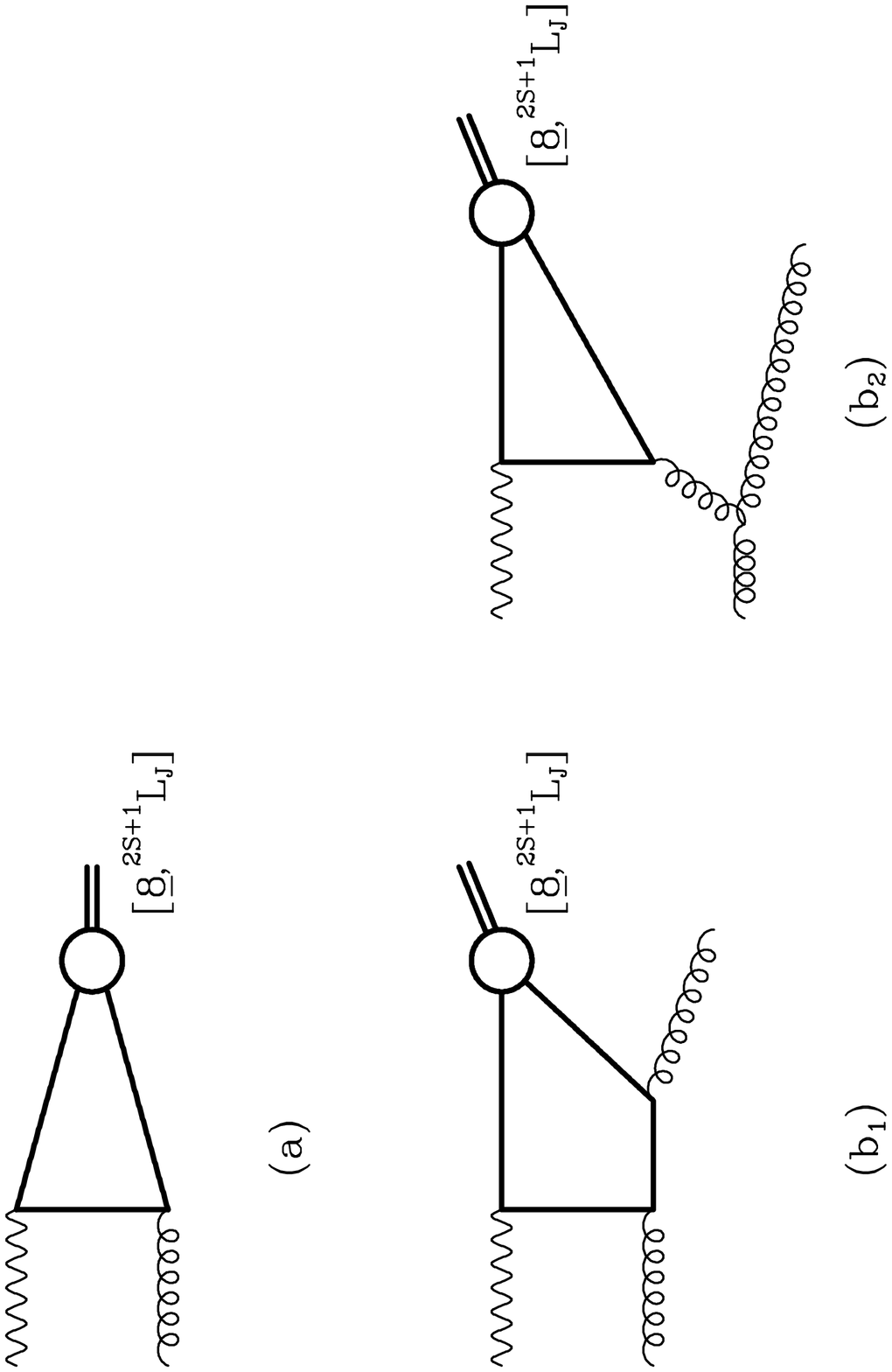}
\end{picture}

\vspace*{3cm}

\fcaption{\label{fig_2} 
  Generic diagrams for $J/\psi$ photoproduction via color-octet
  channels: (a) leading color-octet contributions; (b) color-octet
  contributions to inelastic $J\!/\!\psi$ production.}

\end{figure}

\vspace*{-0.5pt}
\noindent
Color-octet configurations are produced at leading order in
$\mbox{$\alpha_{\mbox{\scriptsize s}}$}$ through the $2\to 1$ parton
processes\cite{CK96,AFM,KLS}
\begin{eqnarray}\label{eq_oc0}
\gamma + g &\! \to \!& c\bar{c}\, [\mbox{$\underline{8}$},{}^1S_{0}]
\nonumber \\
\gamma + g &\! \to \!& c\bar{c}\, [\mbox{$\underline{8}$},{}^3P_{0,2}]
\end{eqnarray}
shown in Fig.\ref{fig_2}a. Due to kinematical constraints, the leading
color-octet terms will only contribute to the upper endpoint of the
$J/\psi$ energy spectrum, $z\approx 1$ and $p_\perp\approx 0$,
$p_\perp$ being the $J/\psi$ transverse momentum.  Color-octet
configurations which contribute to inelastic $J/\psi$ photoproduction
$z \le 0.9$ and $p_\perp \ge 1$~GeV are produced through the
subprocesses\cite{CK96,KLS}
\begin{eqnarray}\label{eq_oc2}
\gamma + g &\! \to \!& c\bar{c}\, [\mbox{$\underline{8}$},{}^1S_{0}] 
  + g \nonumber \\
\gamma + g &\! \to \!& c\bar{c}\, [\mbox{$\underline{8}$},{}^3S_{1}] 
  + g \nonumber \\
\gamma + g &\! \to \!& c\bar{c}\, [\mbox{$\underline{8}$},{}^3P_{0,1,2}] + g 
\end{eqnarray}
as shown in Fig.\ref{fig_2}b. Light-quark initiated contributions are
strongly suppressed at \mbox{HERA} energies and can safely be
neglected.

The transition of the color-octet $c\bar{c} \,
[\mbox{$\underline{8}$},{}^{2S+1}L_{J}]$ pair into a physical $J/\psi$
state through the emission of non-perturbative gluons is described by
the long-distance matrix elements $\langle {\cal{O}}^{J/\psi} \,
[\mbox{$\underline{8}$},{}^{2S+1}L_{J}] \rangle$.  They have to be
obtained from lattice si\-mu\-la\-ti\-ons\cite{BSK96} or measured
directly in some production process.  According to the velocity
scaling rules of NRQCD, the color-octet matrix elements associated
with $S$-wave quarkonia should be suppressed by a factor of $v^4$
compared to the leading color-singlet matrix element.\footnote{In the
  case of $P$-wave quarkonia, color-singlet and color-octet matrix
  elements contribute at the same order in $v$.\cite{BBL92}
  Photoproduction of $P$-wave states is, however, suppressed compared
  with $J/\psi$ states, by two orders of magnitude at
  \mbox{HERA}.\cite{MA,CKHERA}} Color-octet contributions to $J/\psi$
photoproduction can thus become important only if the corresponding
short-distance cross sections are enhanced as compared to the
color-singlet process. Color-octet matrix elements have been fitted to
prompt $J/\psi$ data from \mbox{CDF}\cite{CDF} and found to be
\mbox{${\cal O}(10^{-2}$~GeV$^3)$}, consistent with the NRQCD velocity
scaling rules.\cite{TEV1,CL96} Meanwhile, fit values for color-octet
matrix elements have also been obtained from analyses of quarkonium
production in hadronic collisions at fixed-target
experiments\cite{BR}, $J/\psi$ photoproduction at the elastic
peak\cite{AFM} and $J/\psi$ production in $B$ decays\cite{KLS2}.
The results seem to indicate that the values for the color-octet
matrix elements extracted from the Tevatron data at moderate $p_\perp$
are too large; they should however be considered with some caution
since significant higher-twist corrections are expected to contribute
in the small-$p_\perp$ region probed at fixed target experiments and
in elastic $J/\psi$ photoproduction. Moreover, the comparison between
the different analyses is rendered difficult by the fact that the
color-octet matrix elements can in general only be extracted in
certain linear combinations which depend on the reaction under
consideration, see Sec.4.

\vspace*{1pt}\textlineskip      
\section{$J/\psi$ photoproduction at HERA}
\vspace*{-0.5pt}
\noindent
The production of $J/\psi$ particles in high energy $ep$ collisions at
\mbox{HERA} is dominated by photoproduction events where the electron
is scattered by a small angle producing photons of almost zero
virtuality. The measurements at \mbox{HERA} provide information on the
dynamics of $J/\psi$ photoproduction in a wide kinematical region,
$30~\mbox{GeV} \; \simlt \; \sqrt{s\hphantom{tk}}\!\!\!\!\!  _{\gamma
  p}\;\simlt\; 200~\mbox{GeV}$, corresponding to initial photon
energies in a fixed-target experiment of $450~\mbox{GeV} \; \simlt \;
E_\gamma \; \simlt \; 20,000~\mbox{GeV}$.  Due to kinematical
constraints, the leading color-octet processes (\ref{eq_oc0})
contribute only to the upper endpoint of the $J/\psi$ energy spectrum,
\mbox{$z\approx 1$} and $p_\perp\approx 0$. The color-singlet and
color-octet predictions (\ref{eq_oc0}) have been compared to
experimental data\cite{H1} obtained in the region $z\ge 0.95$ and
$p_\perp \le 1$~GeV.\cite{CK96} Since the fac\-to\-ri\-za\-tion
approach cannot be used to describe the exclusive elastic channel
$\gamma + p \to J/\psi + p$, elastic contributions had been subtracted
from the data sample.  It was shown that the large cross section
predicted by using color-octet matrix elements as extracted from the
Tevatron fits appears to be in conflict with the experimental data.
It is, however, difficult to put strong upper limits for the octet
terms from a measurement of the total cross section in the region
$z\approx 1$ and $p_\perp\approx 0$ since the overall normalization of
the theoretical prediction depends strongly on the choice for the
charm quark mass and the QCD coupling. Moreover, diffractive
production mechanisms which cannot be calculated within perturbative
QCD might contaminate the region $z\approx 1$ and make it difficult to
extract precise information on the color-octet contributions. Finally,
it has been argued that sizable higher-twist effects are expected to
contribute in the region $p_\perp\; \simlt\; 1$~GeV, which cause the
breakdown of the factorization formula (\ref{eq_fac}).\cite{BFY}

It is therefore more appropriate to study $J/\psi$ photoproduction in
the inelastic region $z \le 0.9$ and $p_\perp \ge 1$~GeV where no
diffractive channels contribute and where the general factorization
formula (\ref{eq_fac}) and perturbative QCD calculations should be
applicable.  Adopting the NRQCD matrix elements as extracted from the
fits to prompt $J/\psi$ data at the Tevatron one finds that
color-octet and color-singlet contributions to the inelastic cross
section are predicted to be of comparable size.\cite{CK96,KLS} The
short-distance factors of the $[\mbox{$\underline{8}$},{}^{1}S_{0}]$
and $[\mbox{$\underline{8}$},{}^{3}P_{0,2}]$ channels are strongly
enhanced as compared to the color-singlet term and partly compensate
the ${\cal{O}}(10^{-2})$ suppression of the corresponding
non-perturbative matrix elements.  In contrast, the contributions from
the $[\mbox{$\underline{8}$},{}^{3}S_{1}]$ and
$[\mbox{$\underline{8}$},{}^{3}P_{1}]$ states are suppressed by more
than one order of magnitude.  Since color-octet and color-singlet
processes contribute at the same order in
$\mbox{$\alpha_{\mbox{\scriptsize s}}$}$, the large size of the
$[\mbox{$\underline{8}$},{}^{1}S_{0}]$ and
$[\mbox{$\underline{8}$},{}^{3}P_{0,2}]$ cross sections could not have
been anticipated from naive power counting and demonstrates the
crucial dynamical role played by the bound state quantum
numbers.\cite{BR83} As for the total inelastic cross section, the
linear combination of the color-octet matrix elements $\langle
{\cal{O}}^{J/\psi} \, [\mbox{$\underline{8}$},{}^{1}S_{0}] \rangle$
and $\langle {\cal{O}}^{J/\psi} \,
[\mbox{$\underline{8}$},{}^{3}P_{0}] \rangle$ that is probed at
\mbox{HERA} is almost identical to that extracted from the Tevatron
fits at moderate $p_\perp$, independent of
$\sqrt{s\hphantom{tk}}\!\!\!\!\!  _{\gamma p}$.\footnote{At leading
  order in $v^2$, the $P$-wave matrix elements are related by
  heavy-quark spin symmetry, $\langle {\cal{O}}^{J/\psi}
  \,[\ueight,{}^{3}P_{J}] \rangle \approx \mbox{$(2J+1)$} \, \langle
  {\cal{O}}^{J/\psi} \, [\ueight,{}^{3}P_{0}] \rangle$.} The Tevatron
results can thus be used to make predictions for color-octet
contributions to the total inelastic $J/\psi$ photoproduction cross
section without further ambiguities.  However, taking into account the
uncertainty due to the value of the charm quark mass and the strong
coupling, the significance of color-octet contributions cannot be
deduced from the analysis of the absolute $J/\psi$ production rates.
In fact, the experimental data can be accounted for by the
color-singlet channel alone, once higher-order QCD corrections are
included and the theoretical uncertainties due to variation of the
charm quark mass and the strong coupling are taken into account, as
demonstrated at the end of this section. The same statement holds true
for the transverse momentum spectrum, since, at small and moderate
$p_\perp$, both color-singlet and color-octet contributions are almost
identical in shape.  At large transverse momenta, $p_\perp \;
\rlap{\lower 3.5 pt \hbox{$\mathchar \sim$}} \raise 1pt \hbox {$>$} \;
10$~GeV, charm quark fragmentation dominates over the photon-gluon
fusion process.\cite{SA94,GRS95} In contrast to what was found at the
Tevatron\cite{PT_TEV}, gluon fragmentation into color-octet states is
suppressed over the whole range of $p_\perp$ in the inelastic region
$z\;\simlt\; 0.9$.\cite{GRS95}

\begin{figure}[ht]

\vspace*{3cm}

\begin{picture}(7,7)
\includegraphics{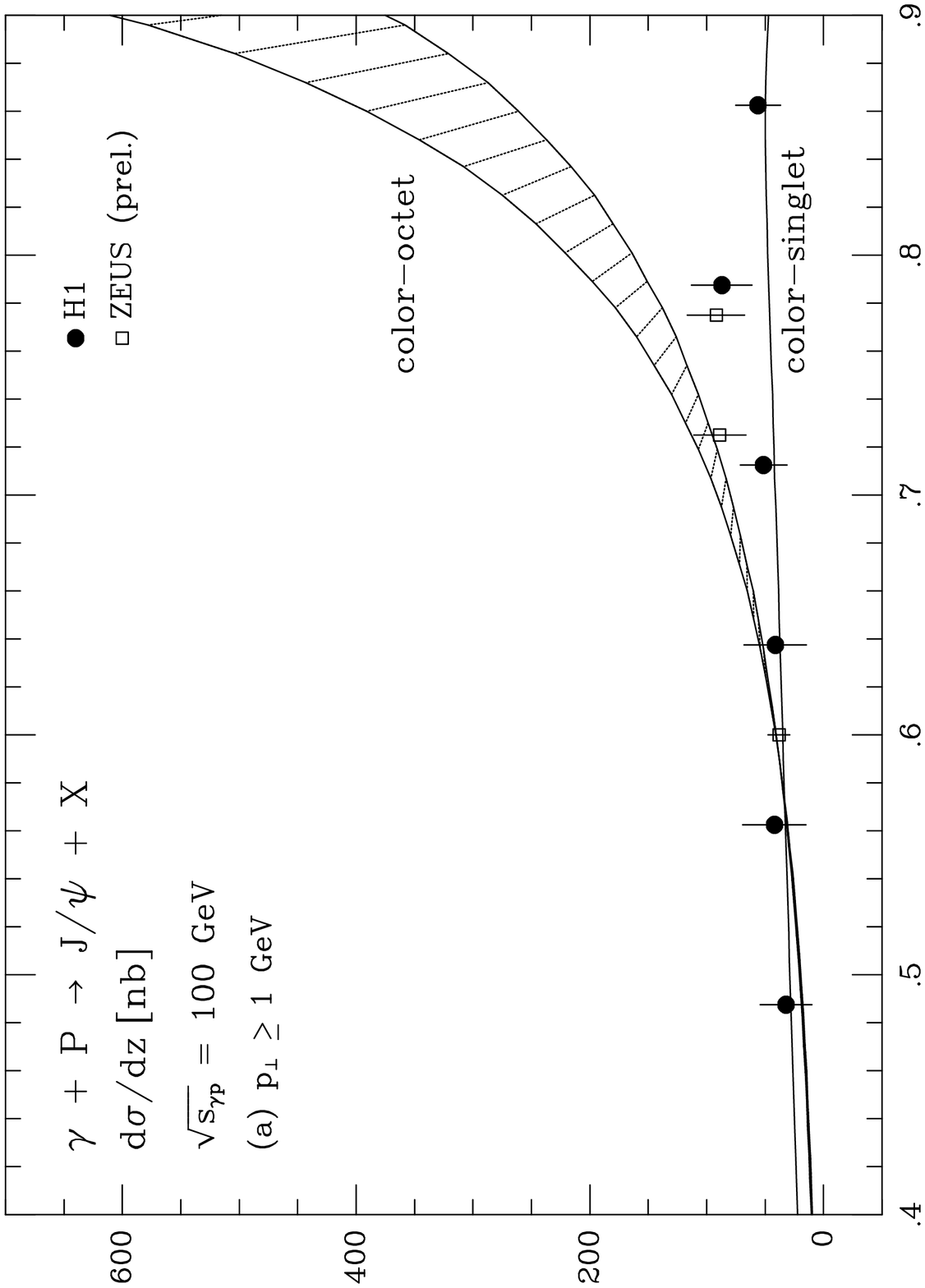}
\end{picture}

\vspace*{5.75cm}

\begin{picture}(7,7)
\includegraphics{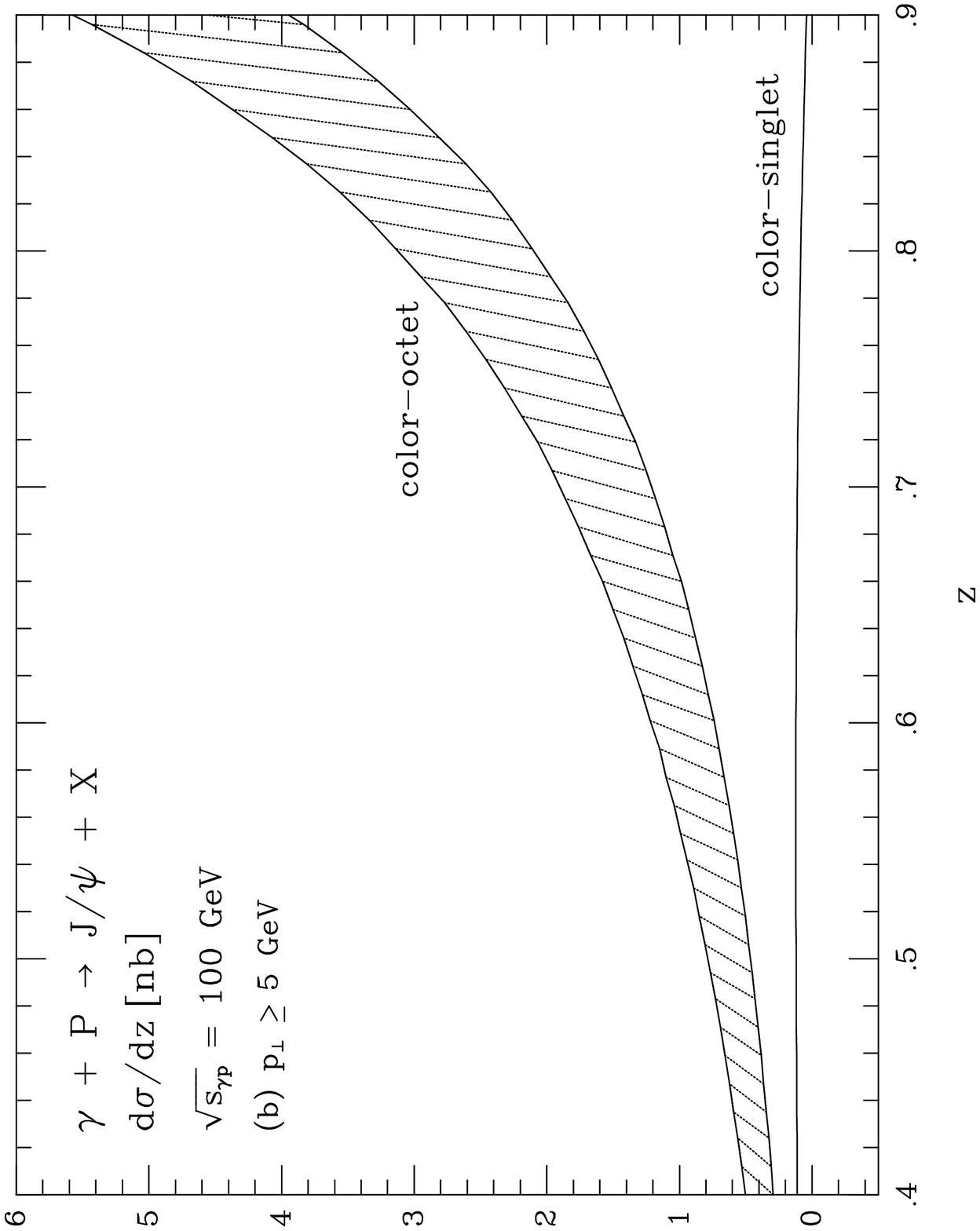}
\end{picture}

\vspace*{3.25cm}

\fcaption{\label{fig_3} Color-singlet and color-octet contributions to
  the $J\!/\!\psi$ energy distribution $\md\sigma/\md{}z$ at the
  photon-proton centre of mass energy $\sqrt{s\hphantom{tk}}\!\!\!\!\!
  _{\gamma p}\,\, = 100$~GeV integrated in the range (a) $p_\perp \ge
  1$~GeV and (b) $p_\perp \ge 5$~GeV compared to experimental
  data\cite{H1,ZEUS}.}

\vspace*{-5mm}

\end{figure}

A distinctive signal for color-octet processes should, however, be
visible in the $J/\psi$ energy distribution
$\mbox{d}\sigma/\mbox{d}{}z$.\cite{CK96} The linear combination of
color-octet matrix elements that is probed by the $J/\psi$ energy
distribution does, however, depend on the value of $z$. Therefore, one
cannot directly use the Tevatron fits but has to allow the individual
color-octet matrix elements to vary in certain ranges, constrained by
the value extracted for the linear combination. It has in fact been
argued that the color-octet matrix element $\langle {\cal{O}}^{J/\psi}
\,[\ueight,{}^{3}P_{0}] \rangle$ could be negative due to the
subtraction of power ultraviolett divergences.\cite{EBpriv} In
contrast, the matrix element $\langle {\cal{O}}^{J/\psi}
\,[\ueight,{}^{1}S_{0}] \rangle$ is free of power divergences and its
value is thus always positive. Accordingly, I have allowed $\langle
{\cal{O}}^{J/\psi} \,[\ueight,{}^{3}P_{0}] \rangle / m_c^2$ to vary in
the range $[-0.01,0.01]$~GeV$^3$ and determined the value of the
matrix element $\langle {\cal{O}}^{J/\psi} \,[\ueight,{}^{1}S_{0}]
\rangle$ from the linear combination extracted at the
Tevatron.\footnote{Note that, given $\langle {\cal{O}}^{J/\psi}
  \,[\ueight,{}^{1}S_{0}] \rangle \;\simlt\; 0.1$~GeV$^3$ as required
  by the velocity scaling rules, a value $\langle {\cal{O}}^{J/\psi}
  \,[\ueight,{}^{3}P_{0}] \rangle / m_c^2 \;\simlt\; -0.01$~GeV$^3$
  would be in contradiction with the Tevatron
  fits.} The result is shown in Fig.\ref{fig_3}(a) where I have plotted
(leading-order) color-singlet and color-octet contributions at a
typical \mbox{HERA} energy of $\sqrt{s\hphantom{tk}} \!\!\!\!\!
_{\gamma p}\,\, = 100$~GeV in the restricted range $p_\perp \ge
1$~GeV, compared to recent experimental data from \mbox{H1}\cite{H1}
and preliminary data from ZEUS\cite{ZEUS}.  The hatched error band
indicates how much the color-octet cross section is altered if
$\langle {\cal{O}}^{J/\psi} \,[\ueight,{}^{3}P_{0}] \rangle / m_c^2$
varies in the range $[-0.01,0.01]$~GeV$^3$, where the lower bound
corresponds to $\langle {\cal{O}}^{J/\psi} \,[\ueight,{}^{3}P_{0}]
\rangle / m_c^2 = -0.01$~GeV$^3$. Since the shape of the distribution
is almost insensitive to higher-order QCD corrections or to the
uncertainty induced by the choice for $m_c$ and
$\mbox{$\alpha_{\mbox{\scriptsize s}}$}$, the analysis of the $J/\psi$
energy spectrum $\mbox{d}\sigma/\mbox{d}{}z$ should provide a clean
test for the underlying production mechanism.  From Fig.\ref{fig_3}
one can conclude that the shape predicted by the color-octet
contributions is not supported by the experimental data.  The
discrepancy with the data can only be removed when reducing the
relative weight of the color-octet contributions by at least a factor
of five.\cite{CK96} Let me emphasize that the rise of the cross
section towards large $z$ predicted by the color-octet mechanism is
not sensitive to the small-$p_\perp$ region and thus not affected by
the collinear divergences which show up at the endpoint $z=1$ and
$p_\perp=0$. This is demonstrated in Fig.\ref{fig_3}(b) where I show
color-singlet and color-octet contributions to the $J/\psi$ energy
distribution for $p_\perp > 5$~GeV. It will be very interesting to
compare these predictions with data to be expected in the future at
\mbox{HERA}. Let me finally mention that the shape of the $J/\psi$
energy distribution could be influenced by the emission of soft gluons
from the intermediate color-octet state.\cite{BR} While this effect,
which cannot be predicted within the NRQCD factorization approach,
might be significant at the elastic peak, it is by no means clear if
and in which way it could affect the inelastic region $z \;\simlt\;
0.9$ and $p_\perp\;\simgt\;1$~GeV.  In fact, if soft gluon emission
were important, it should also result in a feed-down of the leading
color-octet contributions (\ref{eq_oc0}) into the inelastic domain,
thereby increasing the discrepancy between the color-octet cross
section and the data in the large-$z$ region.

\begin{figure}[htbp]

\vspace*{3cm}

\begin{picture}(7,7)
\includegraphics{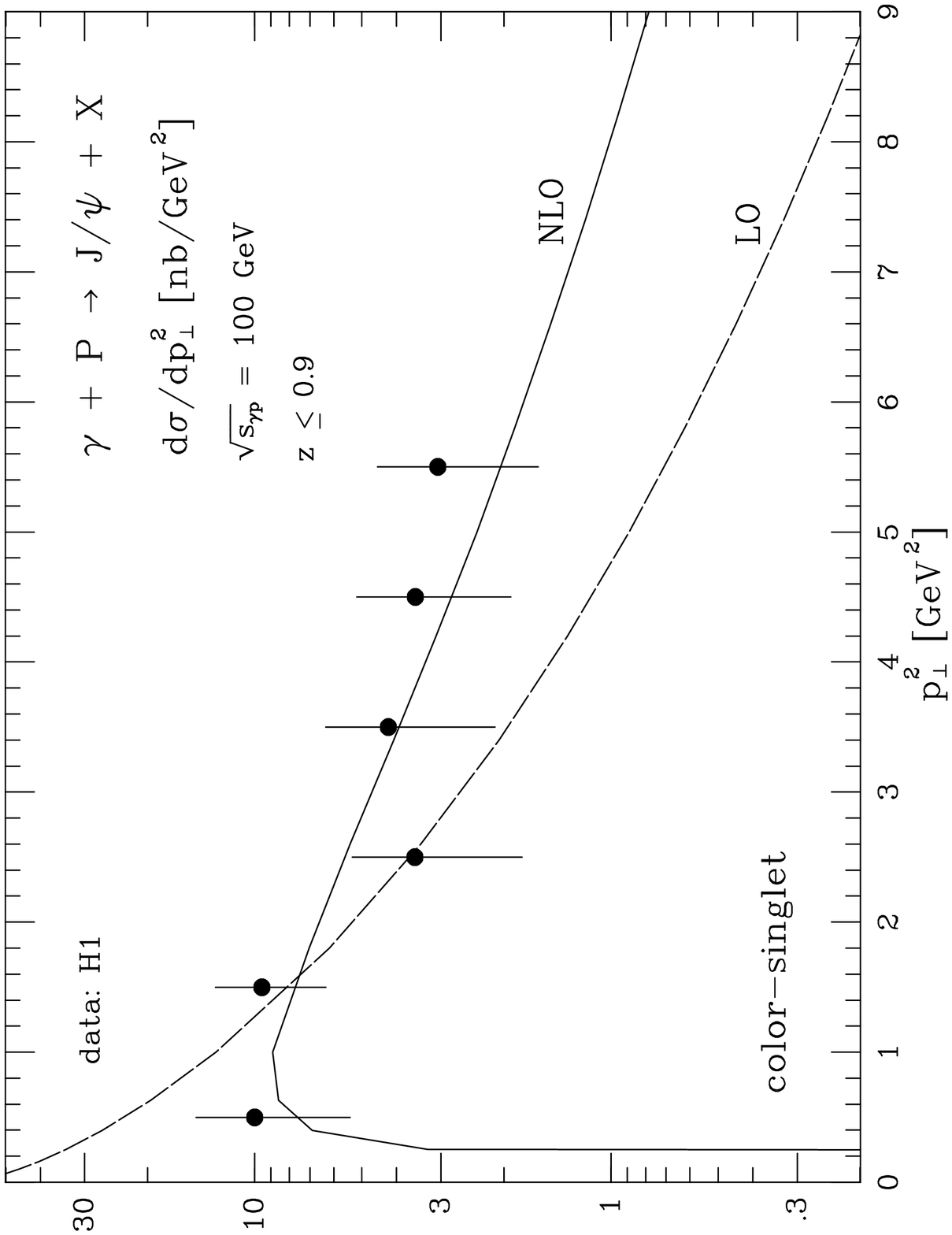}
\end{picture}

\vspace*{3.25cm}

\fcaption{\label{fig_4} LO and NLO color-singlet prediction for the
  $J\!/\!\psi$ transverse momentum spectrum $\md\sigma/\md{}p_\perp^2$
  at the photon-proton centre of mass energy
  $\sqrt{s\hphantom{tk}}\!\!\!\!\!  _{\gamma p}\,\, = 100$~GeV
  integrated in the range $z \le 0.9$ compared to experimental
  data\cite{H1}.}

\end{figure}

\begin{figure}[htbp]

\vspace*{3cm}

\begin{picture}(7,7)
\includegraphics{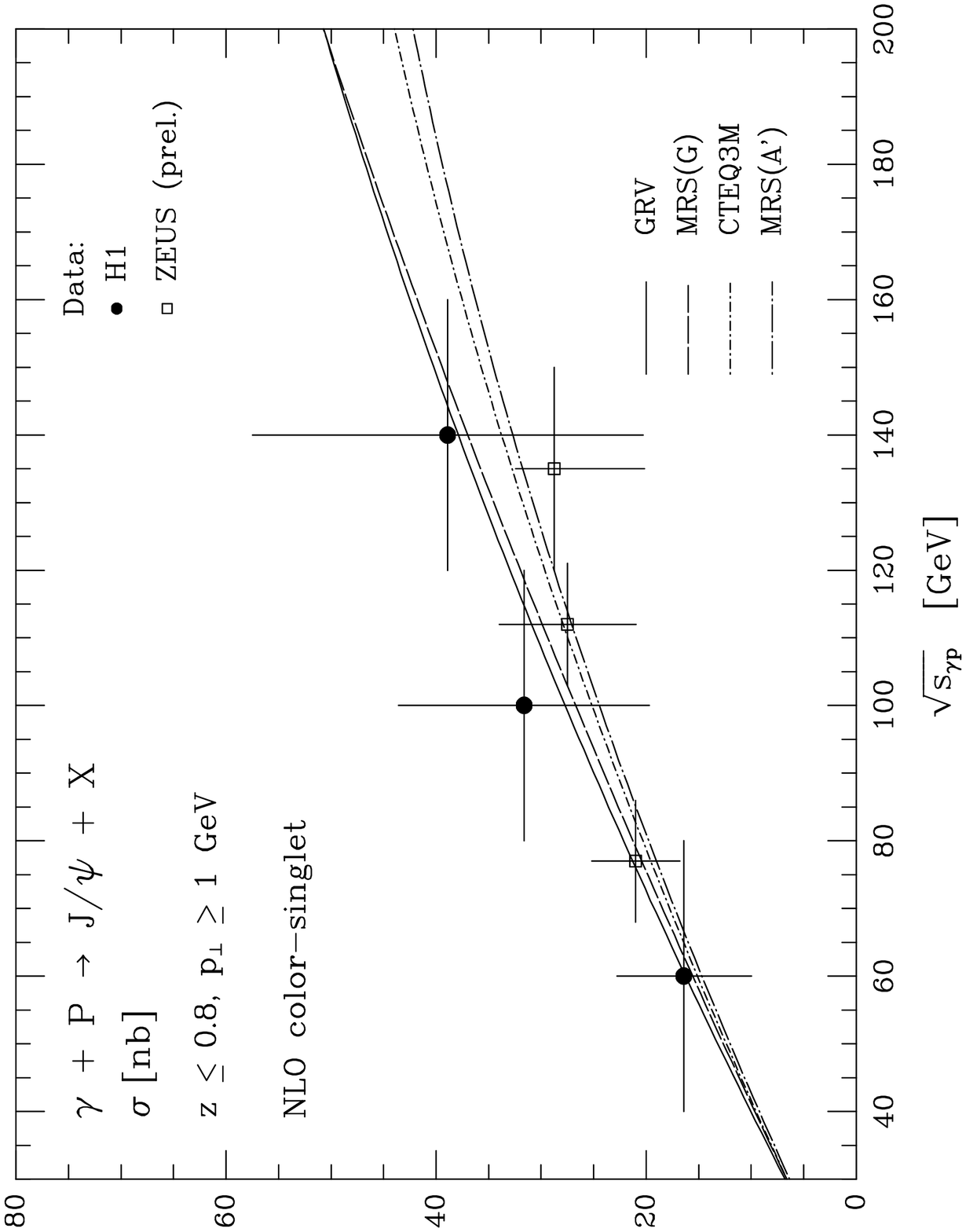}
\end{picture}

\vspace*{3.5cm}

\fcaption{\label{fig_5} NLO color-singlet prediction for the total
  inelastic $J\!/\!\psi$ photoproduction cross section as a function
  of the photon-proton energy for different
  parametrizations\cite{pdfs} of the parton distribution in the proton
  compared to experimental data\cite{H1,ZEUS}.}

\end{figure}

For the remainder of this section, I will demonstrate that the
experimental results on differential distributions and total cross
sections are well accounted for by the color-singlet channel alone
including higher-order QCD corrections.  This can e.g.\ be inferred
from Fig.\ref{fig_4} where I compare the NLO color-singlet prediction
for the $J/\psi$ transverse momentum distribution\cite{MK95} with
recent results from \mbox{H1}\cite{H1}.
Note that the inclusion of higher-order QCD corrections is crucial to
describe the shape of the $p_\perp$ distribution. However, a detailed
analysis of the transverse momentum spectrum reveals that the
fixed-order perturbative QCD calculation is not under proper control
in the limit $p_\perp \to 0$, Fig.\ref{fig_4}.  No reliable prediction
can be made in the small-$p_\perp$ domain without resummation of large
logarithmic corrections caused by multiple gluon emission.  If the
region $p_\perp \le 1$~GeV is excluded from the analysis, the
next-to-leading order color-singlet prediction accounts for the energy
dependence of the cross section and for the overall normalization,
Fig.~\ref{fig_5}. The sensitivity of the prediction to the
small-$x$ behaviour of the gluon distribution is however not very
distinctive, since the average momentum fraction of the partons
$<\!x\!>$ is shifted to larger values when excluding the
small-$p_\perp$ region.

\vspace*{1pt}\textlineskip      
\section{Conclusion}
\vspace*{-0.5pt}
\noindent
I have discussed color-singlet and color-octet contributions to the
production of $J/\psi$ particles in photon-proton collisions,
including higher-order QCD corrections to the color-singlet channel.
A comparison with photoproduction data obtained at fixed-target
experiments\cite{MK95} and the $ep$ collider \mbox{HERA} reveals that
the $J/\psi$ energy spectrum and the slope of the transverse momentum
distribution are adequately accounted for by the next-to-leading order
color-singlet prediction in the inelastic region $p_\perp
\;\simgt\;1$~GeV and $z\;\simlt\;0.9$. Taking into account the
uncertainty due to variation of the charm quark mass and the strong
coupling, one can conclude that the normalization too appears to be
under semi-quantitative control. Higher-twist effects\cite{HT} must be
included to improve the quality of the theoretical analysis further.
Distinctive signatures for color-octet processes should be visible in
the shape of the $J/\psi$ energy distribution. However, these
predictions appear at variance with recent experimental data obtained
at \mbox{HERA} indicating that the values of the color-octet matrix
elements $\langle {\cal{O}}^{J\!/\!\psi} \, [\ueight,{}^{1}S_{0}]
\rangle$ and $ \langle {\cal{O}}^{J\!/\!\psi} \, [\ueight,{}^{3}P_{0}]
\rangle$ are considerably smaller than suggested by the fits to
Tevatron data at moderate $p_\perp$. Support is added to this result
by recent analyses on $J/\psi$ production in hadronic collisions at
fixed-target energies\cite{BR} and $B$ decays\cite{KLS2}.  Clearly,
much more effort, both theoretical and experimental, is needed to
establish the phenomenological significance of color-octet
contributions to $J/\psi$ production and to proof the applicability of
the NRQCD factorization approach to charmonium production in hadronic
collisions at moderate transverse momentum.

\nonumsection{Acknowledgements}
\noindent
I wish to thank Martin Beneke, Eric Braaten, Matteo Cacciari, Sean Fleming 
and Arthur Hebecker for useful discussions. 

\nonumsection{References}

\end{document}